\documentclass[showpacs,preprintnumbers,amsmath,amssymb,twocolumn,superscriptaddress,prb]{revtex4}
\usepackage{graphicx}
\usepackage{amsfonts}
\usepackage{amsmath, amsthm, amssymb}
\usepackage{dsfont}
\usepackage{times}

\newcommand{\ve}[1]{\mathbf{#1}}
\newcommand{\vk}{\ve{k}} 

\newcommand{\e}[1]{\mathrm{e}^{#1}}

\newcommand{\etal}{\emph{et al.}}
\def\i{\mathrm{i}}

\begin{document}
\title[Theory of Andreev reflection in junctions with iron-based High-$T_c$ superconductors]{Theory of Andreev reflection in junctions with iron-based High-$T_c$ superconductors}
\author{Jacob Linder}
\affiliation{Department of Physics, Norwegian University of
Science and Technology, N-7491 Trondheim, Norway}
\author{Asle Sudb{\o}}
\affiliation{Department of Physics, Norwegian University of
Science and Technology, N-7491 Trondheim, Norway}

\date{Received \today}
\begin{abstract}
\noindent We construct a theory for low-energy quantum transport in normal$\mid$superconductor junctions involving the recently 
discovered iron-based high-$T_c$ superconductors. We properly take into account both Andreev bound surface states and the 
complex Fermi surface topology in our approach, and investigate the signatures of the possible order parameter symmetries 
for the FeAs-lattice. Our results could be helpful in determining the symmetry of the superconducting state in the 
iron-pnicitide superconductors.
\end{abstract}
\pacs{74.20.Rp, 74.50.+r, 74.70.Dd}

\maketitle

\textit{Introduction}. Very recently, a family of iron-based superconductors with high transition temperatures was discovered, 
with a concomitant avalanche of both experimental and theoretical activity. \cite{kamihara_jacs_08, wen_arxiv_08, chen_arxiv_08, ren_arxiv_08, mazin_arxiv_08, liu_arxiv_08, cvetkovic_arxiv_08, raghu_prb_08, drew_arxiv_08, chen_nature_08, shan_epl_08, stanev_arxiv_08}.
The highest $T_c$ measured so far in this class of materials is 55 K, and many experimental reports indicate signatures of unconventional superconducting pairing. However, it remains to be clarified what the exact symmetry is for both the orbital- and spin-part of the Cooper pair wavefunction -- there has for instance been reports of both nodal \cite{shan_epl_08} and fully gapped \cite{chen_nature_08} order parameters (OPs) in the literature up to now. 
\par
Probing the low-energy quantum transport properties of superconducting materials has proven itself as a highly useful tool to access information about the symmetry of the superconducting OP. \cite{deutscher_rmp_05} The conductance spectra of normal$\mid$superconductor (N$\mid$S) junctions often contains important and clear signatures of the orbital structure of the OP. For instance, when the OP contains nodes in the tunneling direction with a sign-change across the nodes on each side of the Fermi surface, the conductance will display a large zero-bias conductance peak (ZBCP) due to the presence of Andreev surface bound states. \cite{hu_prl_94}
\par
Two recent studies \cite{shan_epl_08, chen_nature_08} utilized the method of point-contact spectroscopy in order to study the symmetry of the superconducting OP in LaO$_{0.9}$F$_{0.1-\delta}$FeAs and SmFeAsO$_{0.85}$F$_{0.15}$, respectively. The findings were in stark contrast. Namely, 
the large ZBCP found in LaO$_{0.9}$F$_{0.1-\delta}$FeAs gave evidence of a nodal order parameter, while the data of SmFeAsO$_{0.85}$F$_{0.15}$ 
clearly indicated a nodeless OP. In both of these studies, the Blonder-Tinkham-Klapwijk \cite{blonder_prb_82} (BTK) framework was used to 
analyze the data theoretically, using the extension to anisotropic pairing by Tanaka and Kashiwaya \cite{tanaka_prl_95}. In this model, 
one considers a cyndrical or spherical Fermi surface with a free-electron dispersion relation, which does not account for the non-trivial 
multiband Fermi surface topology and dispersion relation in the iron-pnicitides. One might argue that the extended BTK model nevertheless 
may suffice to describe the transport properties of these materials qualitatively, but this statement clearly warrants a detailed investigation. 
\par
In this Rapid Communication, we construct a theory of low-energy quantum transport properties of the iron-based high-$T_c$ superconductors 
by considering a N$\mid$S junction relevant for point-contact spectroscopy and scanning-tunneling-microscopy measurements. In doing so, we 
model fairly accurately the Fermi surface topology and the associated quasiparticle dispersions, in order to see how this affects the results 
as compared with the usual BTK-paradigm. We consider several possible OP symmetries which may be realized in the iron-pnicitides.
\begin{figure}[b!]
\centering
\resizebox{0.40\textwidth}{!}{
\includegraphics{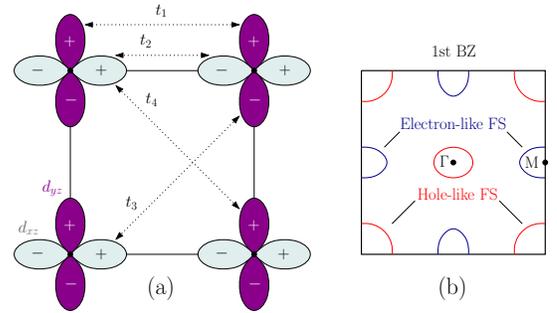}}
\caption{(Color online) (a) Illustration of the two-dimensional FeAs-plane with the $d_{xz}$- and $d_{yz}$-orbitals and hopping between them, as proposed in Ref. \cite{raghu_prb_08}. (b) Sketch of the Fermi surface topology for the long-lived quasiparticle excitations in a minimal two-band model (see main text for parameter values).}
\label{fig:model}
\end{figure}
\par
\textit{Theory and results}. We adopt the minimal two-band model derived in Ref. \cite{raghu_prb_08} (Fig. \ref{fig:model}a) , in which the normal-state Hamiltonian reads
\begin{align}
H_N = \sum_{\vk\sigma} \phi_{\vk\sigma}^\dag\begin{pmatrix}
\epsilon_{\vk x} -\mu & \epsilon_{\vk xy}\\
\epsilon_{\vk xy} & \epsilon_{\vk y} - \mu\\
\end{pmatrix} \phi_{\vk\sigma},
\end{align}
where the fermion basis $\phi_{\vk\sigma} = [d_{\vk x \sigma}, d_{\vk y \sigma}]^\text{T}$ contains the annihilation operators for electrons in the $d_{xz}$- and $d_{yz}$-orbitals with spin $\sigma$ and wavevector $\vk$, respectively. We have also defined $\epsilon_{\vk x} = -2t_1c_x-2t_2c_y - 4t_3c_xc_y$, $\epsilon_{\vk xy} = -4t_4s_xs_y,$ $\epsilon_{\vk y} = -2t_2c_x-2t_1c_y - 4t_3c_xc_y,$
with $c_j = \cos(k_ja), s_j = \sin(k_ja)$, $j=x,y$, and $a$ is the lattice constant.
By diagonalizing the above Hamiltonian, one obtains
\begin{align}\label{eq:normaldispersion}
H_N &= \sum_{\vk\sigma} \tilde{\phi}_{\vk\sigma}^\dag \text{diag}\{\Omega_{\vk}^+, \Omega_{\vk}^-\} \tilde{\phi}_{\vk\sigma},\notag\\
\Omega_{\vk}^\pm &= (\epsilon_{\vk x}+\epsilon_{\vk y})/2 -\mu \pm \sqrt{(\epsilon_{\vk x}-\epsilon_{\vk y})^2/4 + \epsilon_{\vk xy}^2}
\end{align}
where the new basis $\tilde{\phi}_{\vk\sigma} = [\gamma_{\vk\sigma}^+,\gamma_{\vk\sigma}^-]^\text{T}$ consists of new fermion quasiparticle operators in the bands $+$ and $-$ which are hybrids of the $d_{xz}$- and $d_{yz}$-orbitals. The Fermi surface topology is given by $\Omega_\vk^\pm=0$, and gives an electron-like band $(+)$ and hole-like band $(-)$ shown in Fig. \ref{fig:model}b for the choice $t_1=-1$, $t_2=1.3$, $t_3=t_4=-0.85$, $\mu=1.54$, all measured in units of $|t_1|$. Our choice of parameter set is motivated by the fact that it reproduces the same Fermi surface structure as LDA band structure calculations \cite{xu_arxiv_08}, and was also employed in Ref. \cite{parish_arxiv_08, seo_arxiv_08}. The new fermion operators are related to the old basis $\phi_{\vk\sigma}$ by 
\begin{align}\label{eq:P}
&\zeta_\vk = \epsilon_{\vk xy}/[(\epsilon_{\vk x}-\epsilon_{\vk y})/2+ \sqrt{(\epsilon_{\vk x}-\epsilon_{\vk y})^2/4 + \epsilon_{\vk xy}^2}],\notag\\
&\phi_{\vk\sigma}^\dag P_\vk = \tilde{\phi}_{\vk\sigma}^\dag,\; P_\vk =  (1+\zeta_\vk^2)^{-1/2}\times\begin{pmatrix}
1 & -\zeta_\vk\\
\zeta_\vk & 1 \\
\end{pmatrix}.
\end{align}
We now introduce a superconducting pairing between the long-lived quasiparticles $\gamma_{\vk\sigma}^\lambda$, $\lambda=\pm$, which then automatically accounts for both inter- and intra-band pairing in the original fermion basis $\phi_\vk$:
\begin{align}\label{eq:sc}
H_\text{SC} &= \sum_{\vk\lambda} \Big(\Delta_{\vk}^\lambda (\gamma_{\vk\uparrow}^\lambda)^\dag (\gamma_{-\vk\downarrow}^\lambda)^\dag + \text{h.c.}\Big).
\end{align}
In this way, we may diagonalize the total Hamiltonian $H=H_N+H_\text{SC}$ by introducing a final fermion basis $\eta_{\vk}^\lambda = [c_{\vk\uparrow}^\lambda, c_{-\vk\downarrow}^\lambda]^\text{T}$ describing the quasiparticle excitations in the superconducting state. After discarding unimportant constants, we find that $H = \sum_{\vk\sigma\lambda} \sigma E_\vk^\lambda (c_{\vk\sigma}^\lambda)^\dag c_{\vk\sigma}^\lambda,\; E_{\vk}^\lambda = [(\Omega_\vk^\lambda)^2 + |\Delta_{\vk}^\lambda|^2]^{1/2}.$
This result is formally identical to a two-band superconductor with gaps $\Delta_\vk^\lambda$ and normal-state dispersions $\Omega_\vk^\lambda$, $\lambda=\pm$. The belonging wavefunctions which describe the quasiparticle excitations read
\begin{align}\label{eq:wave}
\Psi_{\vk}^\lambda &= \Big\{ 
[u_\vk^\lambda, v_\vk^\lambda \e{-\i\phi_{\vk}^\lambda}]^\text{T}
\e{\i \lambda \mathbf{k}^\lambda\cdot\mathbf{r}}, [v_\vk^\lambda\e{\i\phi_\vk^\lambda}, u_\vk^\lambda]^\text{T}
\e{-\i\lambda \mathbf{k}^\lambda\cdot\mathbf{r}}\Big\},\notag\\
(u_\vk^\lambda)^2 &= 1 - (v_\vk^\lambda)^2 = \frac{1}{2}(1 + \sqrt{E^2 - |\Delta_\vk^\lambda|^2}/E ),
\end{align}
for quasiparticles with positive excitation energies $E\geq 0$. Here, $\mathbf{k}^\lambda$ denotes the Fermi momentum for band $\lambda$ while $\e{\i\phi_\vk^\lambda} = \Delta_\vk^\lambda/|\Delta_\vk^\lambda|$.
\par
We have now effectively described the superconducting state as a two-band model with gaps $\Delta_\vk^\pm$ and normal-state dispersions $\Omega_\vk^\pm$. This has allowed us to obtain a simple form for the wavefunctions in Eq. (\ref{eq:wave}) that are to be used in the scattering problem below. The trade-off for this advantage, however, is that the $\vk$-dependence of the gap functions $\Delta_\vk^\pm$ in general will become quite complicated. To see this, we may transform Eq. (\ref{eq:sc}) back to the original fermion basis $\phi_\vk$ by means of our expression for $P_\vk$ in Eq. (\ref{eq:P}) to find that:
\begin{align}
H_\text{SC} &= \sum_\vk \Big( \Delta_{\vk x} d_{\vk x\uparrow}^\dag d_{-\vk x\downarrow}^\dag + \Delta_{\vk y} d_{\vk y\uparrow}^\dag d_{-\vk y\downarrow}^\dag \notag\\
&+ \Delta_{\vk xy} (d_{\vk x\uparrow}^\dag d_{-\vk y\downarrow}^\dag - d_{\vk y\uparrow}^\dag d_{-\vk x\downarrow}^\dag) + \text{h.c.}\Big),
\end{align}
where $\Delta_{\vk x}$ and $\Delta_{\vk y}$ are the intra-orbital gaps while $\Delta_{\vk xy}$ is the inter-orbital gap, defined as $\Delta_{\vk x} = (\Delta_\vk^+ + \zeta_\vk^2 \Delta_\vk^-)/\nu_\vk^+,$ $\Delta_{\vk y} = (\Delta_\vk^- + \zeta_\vk^2 \Delta_\vk^+)/\nu_\vk^+,$
$\Delta_{\vk xy} = \zeta_\vk(\Delta_\vk^+ - \Delta_\vk^-)/\nu_\vk^+$, with $\nu_\vk^\pm = (1\pm\zeta_\vk^2).$
We see that the inter-orbital pairing vanishes in the case where $\Delta_\vk^+=\Delta_\vk^-$. However, we emphasise that our model does account for inter-orbital pairing $\Delta_{\vk xy}$, and that $\Delta_{\vk xy}\neq0$ whenever $\Delta_\vk^+\neq\Delta_\vk^-$. 
Assuming spin-singlet and even-frequency pairing, there are three possible $s$-wave symmetries $\{\Delta_0$,  $\Delta_0(c_x+c_y)$, $\Delta_0c_xc_y\}$ and two possible $d$-wave symmetries $\{\Delta_0 (c_x-c_y)$, $\Delta_0s_xs_y\}$ for the superconducting order parameters $\Delta_{\vk x}$ and $\Delta_{\vk y}$ in terms of the square lattice harmonics. The gaps in the $\pm$ quasiparticle hybridized bands are then obtained as $\Delta_\vk^+ = (\Delta_{\vk x} - \zeta_\vk^2\Delta_{\vk y})/\nu_\vk^-$ and $\Delta_\vk^- = (\Delta_{\vk y} - \zeta_\vk^2\Delta_{\vk x})/\nu_\vk^-$. Note that the extended $s$-wave symmetry $\sim c_xc_y$ changes sign on the electron- and hole-Fermi surfaces, similarly to the $s_\pm$-scenario suggested in Ref. \cite{mazin_arxiv_08}. 
\par
We are now in a position to evaluate the conductance of the system. The presence of a Fermi-vector mismatch between the normal and superconducting side of the junction is assumed to be manifested through an effective decrease in the junction transmission. Since the Fermi velocity may be different in the two bands with normal-state dispersions $\Omega_\vk^\pm$, we allow for different barrier parameters $Z^\pm$ in the two bands. For a specified pairing symmetry, there are then four fitting parameters present: the barrier strength $Z^\lambda$ and gap magnitude $\Delta_0^\lambda$ for band $\lambda=\pm$. By generalizing the results of Refs.~\onlinecite{blonder_prb_82, tanaka_prl_95} to a two-band model which also takes into account the non-trivial Fermi surface topology in Fig. \ref{fig:model}a, we obtain the following expression for the normalized tunneling conductance: $G(eV)/G_0 = \sum_{\lambda,k_y}f(k_y) \sigma_S^\lambda(eV)/[2f(k_y) \sigma_N^\lambda]$, where $\sigma_N^\lambda = [1+(Z^\lambda)^2]^{-1}$ and
\begin{align}
\sigma_S^\lambda(eV) &= \Big[\sigma_N^\lambda[1 + \sigma_N^\lambda|\Gamma_+^\lambda(\vk, eV)|^2 + (\sigma_N^\lambda-1)
\notag\\
&\times|\Gamma_+^\lambda(\vk,eV)\Gamma_-^\lambda(\vk,eV)|^2]\Big]/\Big[|1+(\sigma_N^\lambda-1)\notag\\
&\times\Gamma_+^\lambda(\vk,eV)\Gamma_-^\lambda(\vk,eV) \rho^\lambda(\vk)|^2\Big],\notag\\
\Gamma_\pm^\lambda(\vk,eV) &= \frac{eV - \sqrt{(eV)^2 - |\Delta(\pm \lambda k_x,k_y)|^2}}{|\Delta^\lambda(\pm \lambda k_x,k_y)|},\notag\\
 \rho^\lambda(\vk) &= \frac{\Delta^\lambda(-\lambda k_x,k_y) [\Delta^\lambda(\lambda k_x,k_y)]^*}{|\Delta^\lambda(-\lambda k_x,k_y)\Delta^\lambda(\lambda k_x,k_y)|},
\end{align}
where $f(k_y)=\cos(k_ya/2)$ is a weighting function that models the directional dependence of the incoming quasiparticles. 
The strategy is now to sum the conductance over allowed values $k_y\in[-\pi/a,\pi/a]$ for the electron- ($\lambda=1$) and hole-like ($\lambda=-1$) Fermi surfaces, and solve for $k_x$ from Eq. (\ref{eq:normaldispersion}) by $\Omega_\vk^\lambda=0$ for a given $k_y$. In what follows, we choose equal value for the barrier tranparencies $Z^+=Z^-\equiv Z$ and gap magnitudes $\Delta_0^+=\Delta_0^-\equiv \Delta_0$ in the two bands for simplicity, and add a small imaginary number $\delta$ to the quasiparticle energy to model inelastic scattering: $eV\to eV+\i\delta$, $\delta/\Delta_0=10^{-2}$. 
\begin{widetext}
\text{ }\\
\begin{figure}[h!]
\centering
\resizebox{0.99\textwidth}{!}{
\includegraphics{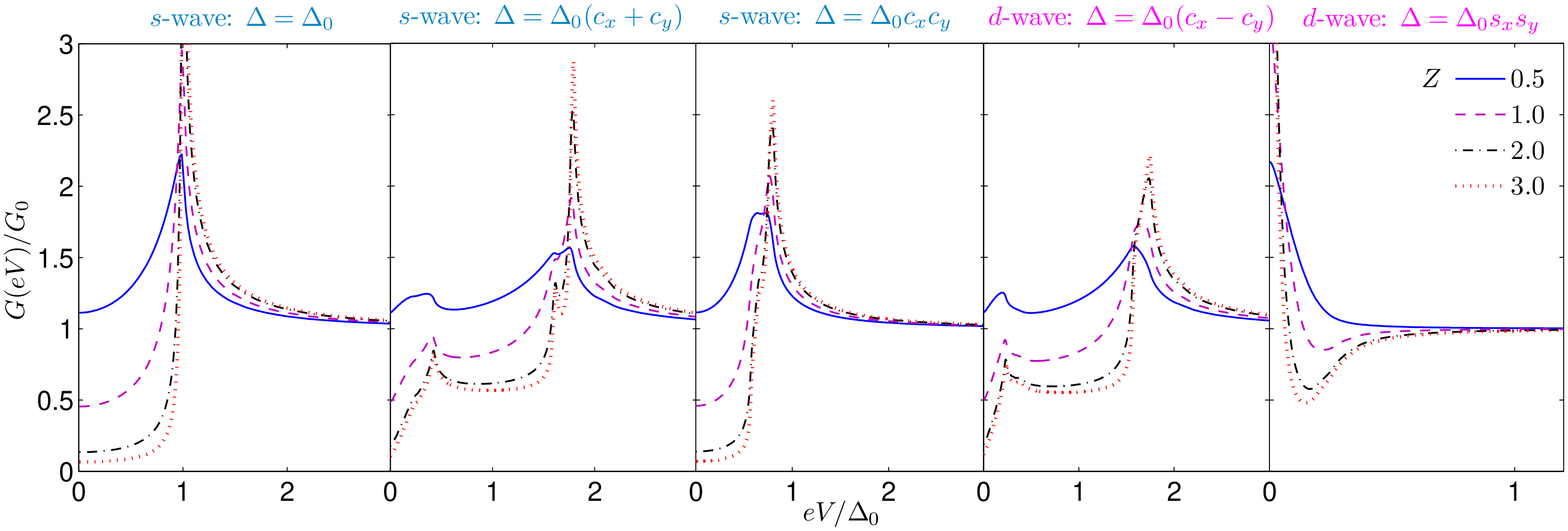}}
\caption{(Color online) Plot of the conductance spectra for tunneling along the (100)-axis in an iron-pnicitide N$\mid$S junction for several possible order parameter symmetries. Only in the $d_{xy}$-wave case $\Delta=\Delta_0s_xs_y$ is there a considerable ZBCP. Note that different scale on the voltage-axis for this case due to the narrowness of the ZBCP. High (low) values of the parameter $Z$ denotes low (high) transmissivity interfaces.}
\label{fig:conductance}
\end{figure}
\end{widetext}
As in Ref. \cite{parish_arxiv_08}, we choose $\Delta_0=0.1$. Clearly, it is possible to study a rich variety of interplays between the two quasiparticle bands in terms of different symmetries for the $d_{xz}$- and $d_{yz}$-orbitals and with different gap magnitudes. Here, however, our main aim is to investigate how the conductance spectra are influenced by the non-trivial Fermi surface topology and dispersion relations, and see how this compares with the cylindrical/spherical Fermi-surface and free-particle dispersion scenario employed in the usual BTK-paradigm. In particular, this is relevant to the interpretation of the point-contact spectroscopy measurements of Refs. \cite{chen_nature_08, shan_epl_08}. There is, however, an important cavaet with regard to which conclusion one may draw with regard to the symmetry of the superconducting OP from the tunneling data of Refs. \cite{chen_nature_08, shan_epl_08}. In these works, polycrystalline samples were used, while the orbital/nodal structure of the OP can only be convincingly probed in single crystal specimens. This is because tunneling into polycrystalline samples may lead to intrinsic averaging effects which distort the contribution from anisotropic OPs. 
\par
In Fig. \ref{fig:conductance}, we plot the conductance for tunneling along the (100)-direction for several OP symmetries. As seen, the $d_{xy}$-wave case stands out from the rest as it features a considerable ZBCP. Comparing with the experimental data of Ref.~\onlinecite{shan_epl_08}, we 
 would conclude that a nodal $d$-wave OP is likely to be realized in LaO$_{0.9}$F$_{0.1-\delta}$FeAs. The results of Ref.~\onlinecite{chen_nature_08} seem to be most consistent with either $s$-wave or extended $s$-wave pairing, as only one gap is seen in the spectra. For the $s$-wave and $d_{xy}$-wave cases, the standard BTK approach appears to suffice in order to qualitatively say something about the OP symmetry. However, the results are quite different from the usual BTK-approach when considering the extended $s$-wave and $d_{x^2-y^2}$-wave symmetries. More specifically, we find satellite features at subgap energies, including sharp peaks. These features most likely pertain to the specific band-structure which we consider here (see Fig. \ref{fig:model}), and are thus not possible to capture within the conventional BTK-treatment with the cylindrical Fermi surface approximation. In fact, the DOS in our minimal two-band model is a highly non-monotonic function of energy and contains two van Hove singularities.\cite{xu_arxiv_08,raghu_prb_08}.
\begin{figure}[b!]
\centering
\resizebox{0.48\textwidth}{!}{
\includegraphics{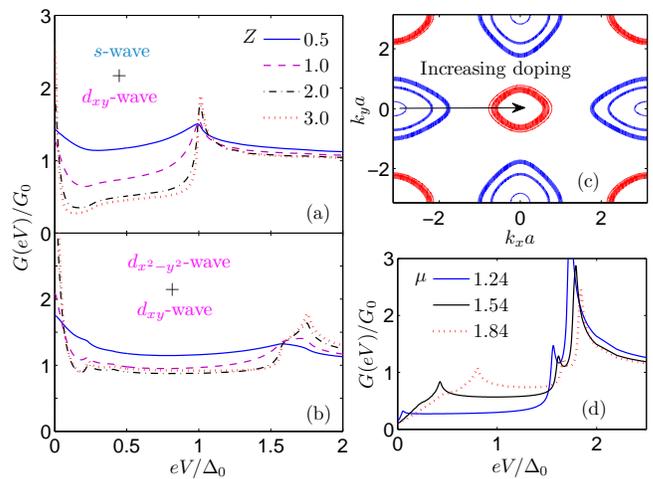}}
\caption{(Color online) (a) and (b): Plot of the conductance spectra for tunneling along the (100)-axis in an iron-pnicitide N$\mid$S junction for the case of one fully gapped OP and one nodal OP. (c) Evolution of the Fermi surface topology for $\mu=\{1.24,1.54,1.84\}$ in the direction of the arrow. (d) Conductance spectra for the $c_x+c_y$ symmetry with $Z=3$ for different doping levels.}
\label{fig:combination}
\end{figure}
Let us also consider the case where there is one fully gapped OP and one nodal OP to see what fingerprints this combination leaves in 
the conductance spectra. In Fig. \ref{fig:combination}, we plot the conductance for the case where $\Delta_{\vk x}$ is fully gapped, 
while $\Delta_{\vk y}$ has a nodal symmetry. For concreteness, we consider $s$-wave + $d_{xy}$-wave pairing and $d_{x^2-y^2}$-wave + $d_{xy}$-wave pairing in Fig. \ref{fig:combination}(a) and (b), respectively. As seen, the nodal OP gives rise to a ZBCP while there are several satellite features in addition to the large coherence peak at the gap edge. The plots are qualitatively similar regardless of whether the fully gapped OP is $s$-wave or $d_{x^2-y^2}$-wave, while the features in the conductance are qualitatively more pronounced in the $s$-wave case due to the better gapping of the Fermi surface. Finally, we consider the evolution of the conductance spectra upon changing the doping level $\mu$. The Fermi surface topology evolves with a change in $\mu$ as shown in Fig. \ref{fig:combination}(c): the electron-pockets increase in size while the hole-pockets decrease in size upon increasing $\mu$. To see how the subgap features obtained in Ref. \ref{fig:conductance} evolve upon modifying $\mu$, consider Fig. \ref{fig:combination}(d) where we consider the $c_x+c_y$ symmetry with $Z=3$. As seen, the satellite features shown in Fig. \ref{fig:conductance} are still present and qualitatively the same, but they are shifted to different bias voltages.
\par
\textit{Summary}. In summary, we have developed a theory for Andreev reflection in the iron-based high-$T_c$ superconductors. Starting with a tight-binding model on a square lattice to model the puckered FeAs planes, we have investigated several order parameter (OP) symmetries and the resulting conductance spectra. Taking fully into account the Fermi surface topology and the quasiparticle dispersion relation, we have investigated scenarios where the symmetry of the superconducting OP in both bands is the same and where it is different, i.e. one is fully gapped and the other is nodal. We find that the standard Blonder-Tinkham-Klapwijk (BTK) formalism should give qualitatively correct results for the case where the OP symmetries on both bands are either isotropic $s$-wave or $d$-wave. However, the results differ considerably for the extended $s$-wave symmetries, as we find satellite features at subgap energies which are absent within the usual BTK treatment. Our results may be useful in the context of analyzing quantum transport data of tunneling in normal$\mid$superconductor junctions involving the iron-pnicitides.

\textit{Acknowledgments.} T. Yokoyama is thanked for useful discussions. J.L. and A.S. were supported by the Research Council of Norway, 
Grants No. 158518/431 and No. 158547/431 (NANOMAT), and Grant No. 167498/V30 (STORFORSK).

\end{document}